\documentclass[prd,twocolumn,superscriptaddress,floatfix,nofootinbib]{revtex4-1}
\pdfoutput=1 %
\usepackage[T1]{fontenc} %
\usepackage{amsmath,amssymb,amsfonts,amsthm}
\usepackage{graphicx}
\usepackage[usenames,dvipsnames]{color}
\usepackage{enumitem}
\usepackage{verbatim}
\usepackage[percent]{overpic} 
\usepackage{rotating}
\usepackage{hyperref}
\usepackage{array}
\usepackage{mathtools}
\usepackage{lmodern}
\usepackage[normalem]{ulem}
\usepackage{braket}
\usepackage{tensor}
\usepackage{dsfont}
\usepackage{bm}
\usepackage{tikz}

\usepackage{mathrsfs}

\usetikzlibrary{decorations.pathmorphing}
\usepackage{CJKutf8}

\newcommand{\kgy}[1]{{\color{blue}\bf [Ken: #1]}}

\newcommand{\pmx}[1]{\begin{pmatrix}#1\end{pmatrix}} %

\newcommand{\lb}{\left}						%
\newcommand{\rb}{\right}					%
\newcommand{\nn}{\nonumber}					%
\newcommand{\bd}{\boldsymbol}				%

\newcommand{\Or}{\mathcal{O}}	\newcommand{\abs}[1]{\left| #1 \right|}

\newcommand{\la}{\lambda}
\newcommand{\rh}{\rho}

\newcommand{\Si}{\Sigma}

\newcommand{\Om}{\Omega}

\usetikzlibrary{positioning}
\usetikzlibrary{calc,through,backgrounds}

\theoremstyle{definition}
\newtheorem{definition}{Definition}[section]
\newtheorem{theorem}{Theorem}[section]

\newcommand{\ts}[1]{ _{\text{#1}} }

\newcommand{\Bigkagikako}[1]{\Big[ #1 \Big]}

\newcommand{\erf}{\text{erf}}
\newcommand{\erfc}{\text{erfc}}

\allowdisplaybreaks

\DeclareMathOperator{\Tr}{Tr}

\newcommand{\id}{\mathds{1}}

\renewcommand{\Im}{\operatorname{Im}}

\begin{document}

\title{Entanglement harvesting of three Unruh-DeWitt detectors}

\author{Diana Mendez-Avalos}
\email[]{dianamendezavalos@gmail.com}
\affiliation{Centre of Nanosciences and Nanotechnology, National Autonomous University of Mexico, Ensenada, Baja California 22860, Mexico}
\author{Laura J. Henderson}
\email[]{l7henderson@uwaterloo.ca}
\affiliation{Department of Physics and Astronomy, University of Waterloo, Waterloo, Ontario, N2L 3G1, Canada}
\affiliation{Centre for Engineered Quantum Systems, School of Mathematics and Physics,
The University of Queensland, St. Lucia, Queensland 4072, Australia}
\author{Kensuke Gallock-Yoshimura}
\email[]{kgallock@uwaterloo.ca}
\affiliation{Department of Physics and Astronomy, University of Waterloo, Waterloo, Ontario, N2L 3G1, Canada}
\author{Robert B. Mann}
\email[]{rbmann@uwaterloo.ca}
\affiliation{Department of Physics and Astronomy, University of Waterloo, Waterloo, Ontario, N2L 3G1, Canada}
\affiliation{Institute for Quantum Computing, University of Waterloo, Waterloo, Ontario, Canada, N2L 3G1}

\begin{abstract}
We analyze a tripartite entanglement harvesting protocol with three Unruh-DeWitt detectors adiabatically interacting with a quantum scalar field. We consider linear, equilateral triangular, and scalene triangular configurations for the detectors.   We find that, under the same parameters,   more entanglement can be extracted in the linear configuration than the equilateral one, consistent with  single instantaneous switching results. No bipartite entanglement is required to harvest tripartite entanglement.
Furthermore,  we find that tripartite entanglement can be harvested even if one   detector is at larger spacelike separations from  the other two than in the corresponding bipartite case.
We also find that for small detector separations bipartite correlations become larger than tripartite ones, leading to an apparent violation of the Coffman-Kundu-Wootters (CKW) inequality. We show that this is not a consequence of our perturbative expansion but that it instead occurs because the harvesting qubits are in a mixed state.

\end{abstract}

\maketitle
\flushbottom

\section{Introduction}
The study of quantum information is inextricably connected to the nature of the quantum vacuum. 
Over the past several decades this subject has been studied from a variety of perspectives, including metrology \cite{Ralph:connectivity} and quantum information \cite{Peres:QuantumInfo,Lamata:Entanglement}, quantum energy teleportation \cite{doi:10.1143/JPSJ.78.034001,Hotta:2011xj}, the AdS/CFT correspondence \cite{Ryu:AdSCFT}, black hole entropy \cite{Solodukhin2011,Brustein2005} and the black hole information paradox \cite{Preskill:1992tc,Mathur:2009hf,almheiri2013black,Braunstein:2009my,Mann:2015luq,Louko2014firewall,Luo:2017}. 
The
vacuum state of any quantum field depends in turn on the structure of spacetime. 
By coupling a quantum field to first-quantized particle detectors, we can gain understanding of the quantum vacuum and its response to the structure of spacetime.

Thanu Padmanabhan, known throughout the physics community as `Paddy', long appreciated the importance of understanding the quantum vacuum. He understood its significance in cosmology
\cite{Joshi:1987zj,Padmanabhan:2004qc,Padmanabhan:2002ji}, in gravity \cite{Padmanabhan:2006hx,Padmanabhan:2006cj}, and in quantum field theory\cite{Sriramkumar:1999nw,Lochan:2016cxt}. He also appreciated the importance of employing particle detector models as a tool that can further our understanding of the vacuum \cite{Sriramkumar:1994pb}.
In the same spirit Paddy had for investigating the quantum vacuum, and using the same tools he employed, we show that even the Minkowski space vacuum has interesting correlation properties that warrant further exploration.

It has been known for quite some time that the  vacuum state of a quantum field has non-local correlations, and that these can be extracted using particle detectors \cite{Valentini1991nonlocalcorr,reznik2003entanglement}. 
Even if the two detectors are not in causal contact throughout the duration of the interaction, the vacuum correlations can be transferred to the  detectors in the form of both mutual information, discord, and entanglement. 
The extracted entanglement can be further distilled into Bell pairs \cite{reznik2005violating}, indicating that in principle the vacuum is a resource for quantum information tasks. 
A considerable amount of research on this phenomenon has since been carried out \cite{Steeg2009,pozas2015harvesting,smith2016topology,Vacuum_Harvesting_experiment2, kukita2017harvesting,Simidzija.Nonperturbative,Simidzija2018no-go,henderson2018harvesting,ng2018AdS,henderson2019entangling,cong2019entanglement,Tjoa2020vaidya,cong2020horizon,Xu:2020pbj,Zhang:2020xvo,Liu:2020jaj,Liu:2021dnl,Hu:2022nxc}, and the process has come to be known as \textit{entanglement harvesting} \cite{salton2015acceleration,Liu:2021dnl}. 
In general it depends on the properties and states of motion of the detectors, and has been investigated for both spacelike and non-spacelike detector separations \cite{olson2012extraction,sabin2012extracting,Brown:Cavity}.

\begin{figure}[t]
  \centering
  \includegraphics[scale=0.3]{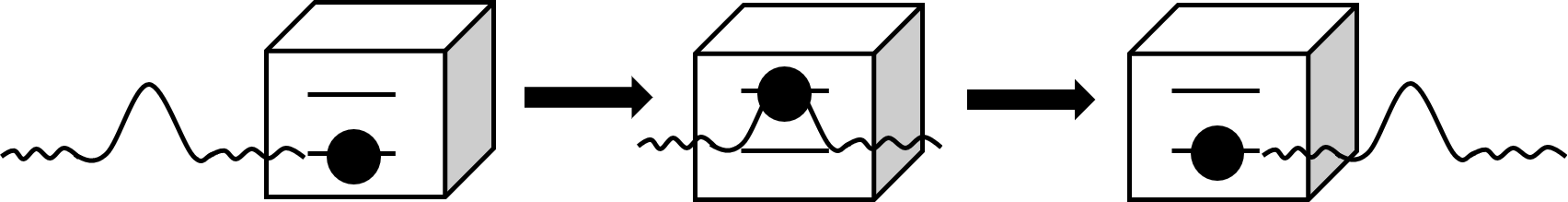}
  \caption{\; UDW detector diagram}
  \label{fig:udwDiagram}
\end{figure}

While a variety of detector models have been considered \cite{Hu2012review,Brown:2012pw,Bruschi:2012rx}, the most popular and well-studied have been  two-level particle detectors, known as Unruh-DeWitt (UDW) detectors\cite{Unruh1979evaporation,DeWitt1979}
that linearly couple to a scalar field. 
These detectors  behave like a particle in a ``box'', which interacts with the field when the ``box'' opens. 
These detectors will be excited by the quantum field when they pass through it (see Fig.~\ref{fig:udwDiagram}). 
Such detectors represent an idealization of atoms responding to an electromagnetic field. 
  Despite its  simplicity, this model captures the essential features of the light-matter interaction \cite{MartinMartinez:2012th,Funai:2018wqq,Funai:2021jpc} whilst remaining illustrative of basic physical principles.  
This has been of particular value in investigating the structure of spacetime \cite{smith2016topology,Ng2017,ng2018AdS,Cong:2021tnk,Dappiaggi:2021wtr}, black holes \cite{Smith2014,Henderson2018, Robbins:Amplification,Kaplanek:2020iay,Tjoa2020vaidya,DeSouzaCampos:2020ddx,deSouzaCampos:2020bnj,Robbins:2021ion,Henderson:2022oyd,Ken.Freefall,KendraBTZ}, gravitational waves \cite{Xu:2020pbj,Gray:2021dfk}, the thermality of de Sitter spacetime \cite{Steeg2009,Huang:2017yjt,Kaplanek:2019vzj,Du:2020cyi}, and the effects of quantum gravity \cite{Foo:2020jmi,Foo:2021fno,Howl:2022oqz, RaureQGwitness, Pitelli.graviton}. Indeed, viewing the situation from a detector perspective and its local interactions with a field is a more operational approach than considering only field interactions, and opens up the  possibility of performing experiments by coupling a physical detector, such as (for example) an atom, qubit, or photon    to the electromagnetic field \cite{Onoe2021}.

Almost all investigations of entanglement harvesting have been concerned with the bipartite case.
However the process of swapping field entanglement can be extended to multipartite entanglement as well, of which much less is known. 
There has been some consideration of tripartite and tetrapartite entanglement in non-inertial frames, \cite{Torres-Arenas2018, Khan2014, Hwang:pra2011,Szypulski:2021} and in inertial frames\cite {Guhne2009}. A finite-duration interaction of $N$ UDW detectors with the scalar vacuum yields a reduced density matrix containing  $N$-partite entanglement between the detectors that can in principle be distilled to that of a $W$-state\cite{SilamReznik:Wstate}.   Gaussian quantum mechanics has been used to show that three harmonic oscillators in a $(1+1)$-dimensional cavity can harvest  genuine tripartite entanglement  even if the detectors remain spacelike separated\cite{Lorek2014tripartite}. 
Curiously, it was easier to harvest tripartite entanglement than bipartite entanglement.  Most recently
extraction of tripartite entanglement using three UDW detectors each instantaneously locally interacting just once with a scalar field was demonstrated \cite{Avalos:2022oxr}, 
whereas a no-go theorem forbids extraction of bipartite entanglement via the same procedure \cite{Simidzija2018no-go}. The harvested tripartite entanglement is of the GHZ-type, and can be maximized by an optimal value of the coupling. 

Here we extend the previous study to include interactions that are not instantaneous, but rather last for a (finite) duration of time. 
Such an interaction allows detectors to harvest entanglement even when they are causally disconnected, whereas the instantaneously interacting detectors in Ref.~\cite{Avalos:2022oxr} require signalling from one detector to the other to gain entanglement. 
Although the elements in the density matrix of three detectors can be complicated in general, one can choose a symmetric spatial configuration for the detectors' positions to simplify   the calculation. 
To this end, we will choose equilateral triangle and linear configurations as simple examples, and then generalize these to a scalene triangle. 
We find that tripartite entanglement can be easily extracted compared to bipartite.  Indeed situations exist  where bipartite entanglement   vanishes but tripartite entanglement can still be harvested.

For small detector separations we find that the total bipartite correlations become larger than the tripartite ones.  This implies an  apparent violation of the Coffman-Kundu-Wootters (CKW) inequality, which describes the monogamy of tripartite entanglement.
We show that this is not a consequence of our perturbative expansion but that it instead occurs because the harvesting qubits are in a mixed state, and provide  in an Appendix an explicit non-perturbative example of of a density matrix for which this is the case.
 
 The outline of our paper is as follows. First, we describe the mathematical method for the Unruh-DeWitt model in the context of our three detector system in Sec.~\ref{sec:UDW}. We then  explore three different detector configurations: equilateral triangular (Sec.~\ref{sec:equilateral}), linear (Sec.~\ref{sec:linear}), and scalene triangular (Sec.~\ref{sec:scalene}). The scalene triangle configuration generalizes of the other two. We summarize our results in Sec.~\ref{sec:Conclusion}, and include two Appendices that provide details of our calculations.

\section{The Unruh-DeWitt model}\label{sec:UDW}

The UDW model of a detector $D$ is a two-level quantum system, with ground and excited states denoted by $\Ket{0}_D$ and $\Ket{1}_D$, respectively, and separated by an energy gap of $\Omega_D$. 
Suppose three UDW detectors, $A, B$, and $C$, locally couple to a massless quantum scalar field~$\hat{\phi}(\bd{x},t)$ whose interaction Hamiltonian is 
\begin{equation}
    \label{eqn:UWDInteractionHamiltonian}
    \hat H(t) 
    =
        \sum_{D=A,B,C} 
        \dfrac{d\tau_D}{dt}
        \lambda_D  \chi_D(\tau_D(t))
        \hat \mu_D(\tau_D(t))
        \otimes \hat\phi[x_D(t)]
\end{equation}
in the interaction picture, 
where $\tau_D$ and $t$ are the proper time of detector $D$ and a common time, respectively, $\lambda_D$ is the coupling of detector $D$ to the field, $\chi_D(\tau_D)$ is a switching function, $\hat \mu_D(\tau_D)$ is a monopole moment given by
\begin{align}
    \hat \mu_D(\tau_D)
    &=
        e^{i\Omega_D \tau_D}\ket{1}_D\bra{0}_D
        +
        e^{-i\Omega_D \tau_D}\ket{0}_D\bra{1}_D\,,
\end{align}
and $\hat \phi[x_D(t)]$ is the pullback of the field operator on detector $D$'s trajectory. 
The parameter space of the detectors is rather broad, with three switching functions, three couplings, and three gaps. 

\begin{widetext}
For simplicity, let us assume that the coupling constants are all the same, $\lambda\coloneqq \lambda_D$, and weakly coupled: $\lambda \ll 1$. 
The time evolution of the detectors and field during the interaction is described by the unitary operator, $\hat U$, generated by the interaction Hamiltonian in \eqref{eqn:UWDInteractionHamiltonian}, which is  
    \begin{equation}
        \label{eqn:UnitaryOperator}
        \begin{split}
        \hat U
        &\coloneqq
        \mathcal{T}
        \exp\left(-i\int_{\mathbb{R}} dt\, \hat H(t) \right) 
        =\id + (-i\lambda)\int_{\mathbb{R}} dt\, \hat H(t)
        + \frac{(-i\lambda)^2}{2}
        \int_{\mathbb{R}} dt
        \int_{\mathbb{R}} dt'\,\mathcal{T}\hat H(t)\hat H(t')
        +\mathcal{O}(\lambda^3)\,,
        \end{split}
    \end{equation}
where we have employed the Dyson series expansion, with $\mathcal{T} \hat A(t) \hat B(t') \coloneqq \theta(t-t')\hat A(t)\hat B(t')+\theta(t'-t)\hat B(t') \hat A(t)$  the time-ordering operator  with respect to the common time $t$, and $\theta(t)$ being Heaviside's step function.  

Suppose the detectors are initially prepared 
(as $t\rightarrow-\infty$) 
in the ground state, $\ket{0_A 0_B 0_C}$, and the field is in an appropriately defined vacuum state $\ket{0}$. The initial density matrix is thus $\rho_0=\ket{0_A0_B0_C}\bra{0_A0_B0_C}\otimes \ket{0}\bra{0}$. 
Then the final state of the detectors after the interaction will be
\begin{align}\label{rhoev}
    \rho_{ABC}
    &=
        \Tr_\phi[ \hat U \rho_0 \hat U^\dag ]\,.
\end{align}
In the basis $\{ \ket{0_{A}0_{B}0_{C}}, \ket{0_{A}0_{B}1_{C}}$, $\ket{0_{A}1_{B}0_{C}}, \ket{1_{A}0_{B}0_{C}}$, $\ket{0_{A}1_{B}1_{C}}$, $\ket{1_{A}0_{B}1_{C}}$, $\ket{1_{A}1_{B}0_{C}}, \ket{1_{A}1_{B}1_{C}} \}$ it is straightforward to show \cite{Avalos:2022oxr} that the general structure of the density matrix \eqref{rhoev} is
\begin{align}
\rho_{ABC}
    &=
        \left(
        \begin{array}{cccccccc}
        r_{11} &0 &0 &0 &r_{51}^* &r_{61}^* &r_{71}^* &0  \\
        0 &r_{22} &r_{32}^* &r_{42}^* &0 &0 &0 &r_{82}^*  \\
        0 &r_{32} &r_{33} &r_{43}^* &0 &0 &0 &r_{83}^*  \\
        0 &r_{42} &r_{43} &r_{44} &0 &0 &0 &r_{84}^*  \\
        r_{51} &0 &0 &0 &r_{55} &r_{65}^* &r_{75}^* &0  \\
        r_{61} &0 &0 &0 &r_{65} &r_{66} &r_{76}^* &0  \\
        r_{71} &0 &0 &0 &r_{75} &r_{76} &r_{77} &0  \\
        0 &r_{82} &r_{83} &r_{84} &0 &0 &0 &r_{88} 
        \end{array}
        \right) \label{eq:rhoABC}
\end{align}
to all orders in the coupling, 
where the matrix elements $r_{ij}$ depend on the parameters of the detectors (their gaps and switching functions) as well as their relative states of motion.

To leading order in $\lambda$ the density matrix \eqref{eq:rhoABC} becomes
\begin{align}
 {\rho}_{ABC} = \pmx{
		1-(P_A+P_B+P_C) & 0 & 0 & 0 & X_{BC}^*  & X_{AC}^* & X_{AB}^* & 0 \\
		0 & P_C & C_{BC}^*  & C_{AC}^* & 0 & 0 & 0 & 0 \\
		0 & C_{BC} & P_B & C_{AB}^*  & 0  & 0 & 0 & 0 \\
		0 & C_{AC} & C_{AB}  & P_A  & 0 & 0 & 0 & 0 \\
		X_{BC} & 0 & 0 & 0 & 0 & 0 & 0 & 0 \\
		X_{AC} & 0 & 0 & 0 & 0 & 0 & 0 & 0 \\ 
		X_{AB} & 0 & 0 & 0 & 0 & 0 & 0 & 0 \\
		0 & 0 & 0 & 0 & 0 & 0 & 0 & 0
	} + \Or(\lambda^4)\,,
	\label{eq:rhoABC2ndOrder}
\end{align}
where 
\begin{align}
    P_D 
    &= 
        \lambda^2 \int_{\mathbb{R}}d\tau_D \int_{\mathbb{R}} d\tau_{D}'\, 
        \chi_D(\tau_D) \chi_D(\tau_D') 
        e^{-i\Omega_D(\tau_D-\tau_D')} W(x_D(\tau_D),x_D(\tau_D'))\,,\label{eqn:DetectorProbability} \\
    C_{DD'}
    &= 
        \lambda^2 
        \int_{\mathbb{R}}d\tau_D \int_{\mathbb{R}} d\tau_{D'}'\, 
        \chi_D(\tau_D) \chi_{D'}(\tau_{D'}') 
        e^{-i(\Omega_D \tau_D - \Omega_{D'} \tau_{D'}')} W(x_D(\tau_D),x_{D'}(\tau_{D'}'))\,, \\
    X_{DD'}
    &= 
        -\lambda^2 
        \int_{\mathbb{R}}d\tau_D \int_{\mathbb{R}} d\tau_{D'}'\, 
        \chi_D(\tau_D) \chi_{D'}(\tau_{D'}') 
        e^{i (\Omega_D \tau_D + \Omega_{D'} \tau_{D'}')} \notag \\
        &\quad\times 
        \Bigkagikako{
            \theta(t(\tau_D)-t(\tau_{D'}'))
            W(x_D(\tau_D),x_{D'}(\tau_{D'}'))
            +
            \theta(t(\tau_{D'}')-t(\tau_{D}))
            W(x_{D'}(\tau_{D'}'), x_D(\tau_D))
        }\,.
\end{align}
\end{widetext}
Here $W(x,x')$ is the vacuum Wightman function 
\begin{equation}
    \label{eqn:WightmanFunction}
     W(x,x'):=\bra{0}\phi(x)\phi(x')\ket{0}\,,
\end{equation}
and $C_{D'D} = C_{DD'}^*$ and $D,D'\in\{A,B,C\}$ with $D\ne D'$. 
(Note that if $D=D'$ then $C_{DD} = P_D$.) 
$P_D$ is called transition probability since the reduced density matrix for detector $D$ is given by
\begin{center}
    \begin{math}
        \rho_{D}=\begin{pmatrix}
        1-P_D & 0  \\
        0 & P_D 
        \end{pmatrix}+\mathcal{O}(\lambda^4)
    \end{math}
\end{center}
obtained by tracing the density matrix \eqref{eq:rhoABC2ndOrder}
over the Hilbert spaces of the other two detectors.

After obtaining the density matrix \eqref{eq:rhoABC2ndOrder}, the next step is to extract its multipartite properties, for which several methods exist \cite{Dur2000,Guhne2009,Ou:pitangle,Shi2022,Halder2021}. 
We shall use \textit{the $\pi$-tangle} \cite{Ou2007}, which will provide a lower bound on the tripartite entanglement in the mixed state of a three detector system:  
\begin{equation}
    \label{eqn:PiTangle}
    \pi
    \coloneqq
        \frac{\pi_A+\pi_B+\pi_C}{3}\,,
\end{equation}
where the Negativities of the 3-detector system and its subsystems are respectively
\begin{subequations}
\begin{align}
    \label{eqn:PiA}
    \pi_A &=\mathcal{N}_{A(BC)}^2-\mathcal{N}_{A(B)}^2-\mathcal{N}_{A(C)}^2\,, \\
    \label{eqn:PiB}
    \pi_B &=\mathcal{N}_{B(AC)}^2-\mathcal{N}_{B(A)}^2-\mathcal{N}_{B(C)}^2\,,\\
    \label{eqn:PiC}
    \pi_C &=\mathcal{N}_{C(AB)}^2-\mathcal{N}_{C(B)}^2-\mathcal{N}_{C(A)}^2\,,
\end{align}
\label{eq:SubPi}
\end{subequations}
and 
\begin{align}
    \mathcal{N}_{A(BC)}
    &\coloneqq 
        \dfrac{||\rho_{ABC}^{T_A}||-1}{2}\,,\label{eqn:NegativityTri} \\
    \mathcal{N}_{A(B)} 
    &\coloneqq 
        \dfrac{||(\Tr_C[\rho_{ABC}])^{T_A}||-1}{2}\,,\label{eqn:NegativityBi}
\end{align}
where $||\cdot||$ is the trace norm, and the above definition is cyclic in $A,B,C$. \footnote{The definition of negativity used in \cite{Ou2007} does not include the overall factor of $\frac{1}{2}$. However, we include it to keep consistent with the definition used in \cite{Vidal2002negativity}.} 
The partial transpose $\rho^{T_A}_{ABC}$ over the detector $A$ can be computed as 
\begin{align}
    \label{eqn:PartialTranspose}
    \rho_{ABC}^{T_A} &= \sum_{ijk,i'j'k'}\eta_{ijk}\eta_{i'j'k'}^*(\ket{ijk}\bra{i'j'k'})^{T_A} \nonumber\\
    &=\sum_{ijk,i'j'k'}\eta_{ijk}\eta_{i'j'k'}^*\ket{i'jk}\bra{ij'k'}
\end{align}
over the basis $\ket{ijk}$,
with analogous expressions straightforwardly holding for the other two detectors. 
The partial trace over detector $D$ is computed as 
\begin{equation}
    \label{eqn:PartialTrace}
    \Tr_D[\rho_{ABC}]= {}_{D}\negthinspace\bra{0}\rho_{ABC}\ket{0}_D + {}_D\negthinspace\bra{1}\rho_{ABC}\ket{1}_D \; .
\end{equation}

\begin{figure}[t]
    \centering
    \includegraphics[width=\linewidth]{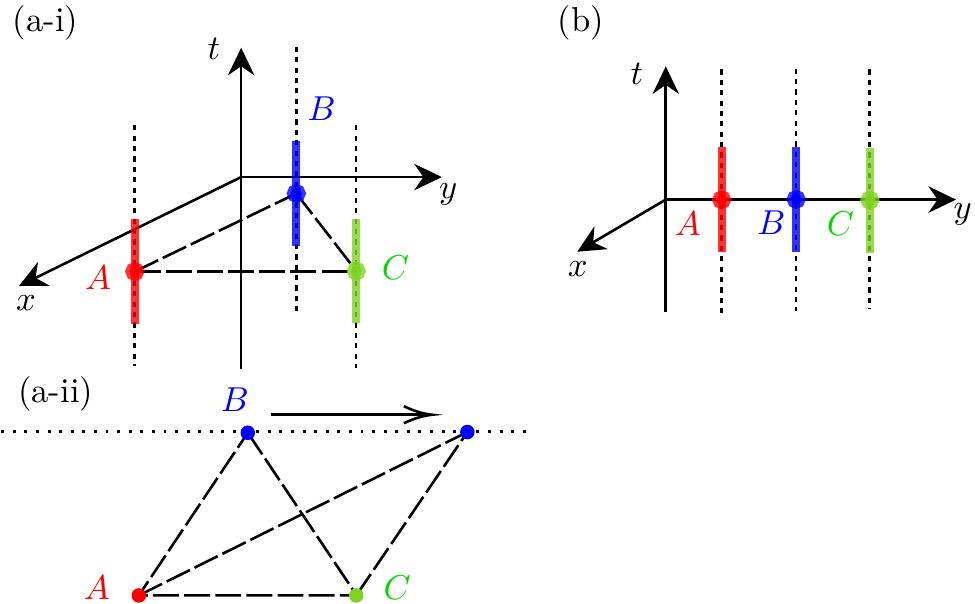}
  \caption{{$\quad$}Three configurations of the 3-detector system; {(a-i)} equilateral triangular, (b) linear, and  {(a-ii)} scalene triangular. Figure
  (a-ii) shows the spatial configuration of the detectors in (a-i) when $t=0$. 
  }
\label{2configs}
\end{figure}

In what follows we shall take all detectors to be identical, having the same energy gap $\Omega$ and switching function $\chi(\tau)$, in order to reduce the complexity of the parameter space.   As a consequence, the three detectors will have identical transition probabilities
\begin{equation}
    P_A = P_B = P_C \equiv P\,.
\end{equation}
In particular, we choose a Gaussian switching function $\chi(\tau)=e^{-\tau^2/2\sigma^2}$, where $\sigma$ is the typical duration of interaction, which allows for the exact calculation of the matrix elements\cite{Smith2017}:
\begin{align}
    P &= \frac{\lambda^2}{4\pi} \lb(e^{-\sigma^2\Omega^2} - \sqrt{\pi}\sigma\Omega\ \erfc\lb(\sigma\Omega\rb)\rb) \\
    C_{DD'} &= \frac{\lambda^2\sigma}{4\sqrt{\pi}L}e^{-L^2/(4\sigma^2)}\bigg[\Im\lb(e^{i\Omega L}\erf\lb(i\frac{L}{2\sigma}+\sigma\Omega\rb)\rb) \nn\\
    &\quad -\sin\lb(\Omega L\rb)\bigg] \in\mathbb{R}\\
    X_{DD'} &= i \frac{\lambda^2\sigma}{4\sqrt{\pi}L} e^{-\sigma^2\Omega^2-L^2/(4\sigma^2)}\lb[1+\erf\lb(i\frac{L}{2\sigma}\rb)\rb]
\end{align}
where
\begin{align*}
    \erf(x) &\coloneqq \frac{2}{\sqrt{\pi}} \int_{0}^{x} dt\, e^{-t^2}\,, \\
    \erfc(x) &\coloneqq 1-\erf(x)
\end{align*}
are the error and complementary error functions respectively and $L\coloneqq\abs{\bd{x}_D-\bd{x}_{D'}}$ is the distance between detectors $D$ and $D'$.

We will consider  three types of detector configurations, shown in
Fig.~\ref{2configs}. In one we place  the detectors at the vertices of an  equilateral triangle, and compute how the entanglement depends on the side length $L$.  In the second configuration we place the detectors in a line of total length $2L$, with the central detector equidistant from the other two, varying the separation of the end detectors from the central one.  Finally, we consider a scalene triangular configuration, in which 
we vary the location of one along a line parallel to that connecting the other two, which remain at fixed separation.

\begin{figure}[t]
    \centering
    \includegraphics[width=\linewidth]{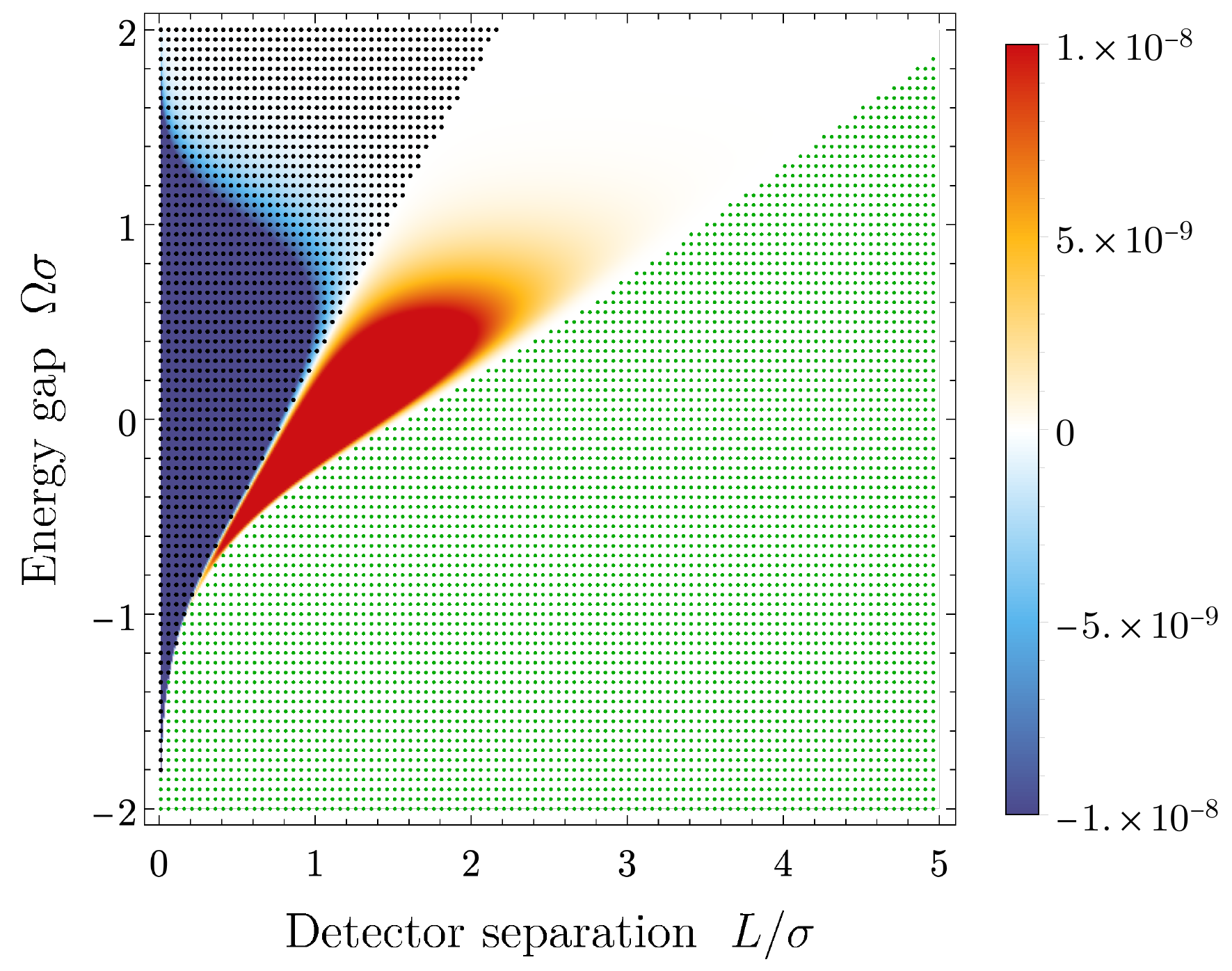}
  \caption{~The $\pi$-tangle as a function of energy gap, $\Omega$, and detector separation, $L$, for the equilateral triangle configuration.  Negative values of $\Omega$ correspond to initially excited detectors.  The coupling constant is set to $\lambda=0.1$.
  The green and black dotted regions represent zero and negative $\pi$-tangle, respectively.}
\label{fig:Triangle}
\end{figure}

\section{Equilateral triangle}\label{sec:equilateral}

The $\pi$-tangle for the  equilateral triangular arrangement is the easiest one to compute since all  distances between the identical detectors are the same and the system is symmetric under permutations of the detectors. As a consequence 
\begin{align} 
    &C_{BC}=C_{AC}=C_{AB} \equiv C\,, \\
    &X_{CB}=X_{CA}=X_{BA} \equiv X\,, \nonumber
\end{align}
and the density matrix \eqref{eq:rhoABC2ndOrder} becomes
\begin{equation}
\label{eqn:ReducedDensMatrixTriangle}
\rho_{ABC}=\begin{pmatrix}
1-3P & 0 & 0 & 0 & X^* & X^* & X^* & 0 \\
0 & P & C & C & 0 & 0 & 0 & 0 \\
0 & C & P & C & 0 & 0 & 0 & 0\\
0 & C & C & P & 0 & 0 & 0 & 0 \\
X & 0 & 0 & 0 & 0  & 0 & 0 & 0 \\
X & 0 & 0 & 0 & 0 & 0 & 0 & 0 \\
X & 0 & 0 & 0 & 0 & 0 & 0 & 0 \\
0 & 0 & 0 & 0 & 0 & 0 & 0 & 0 \\
\end{pmatrix}
+\mathcal{O}(\lambda^4).
\end{equation}
The Negativities \eqref{eqn:NegativityTri} are then  
\begin{align}
    \mathcal{N}_{A(BC)}
    &=
        \max\Bigg[0, \frac{\sqrt{C^2+8\abs{X}^2}}{2} - \frac{C}{2} -P \Bigg]
        +\mathcal{O}(\lambda^4)
\end{align}
with identical results for $ \mathcal{N}_{B(CA)}$ and $ \mathcal{N}_{C(AB)}$, due to  the symmetry of the triangular configuration. 
For the same reason, all bipartite Negativities are the same, and so
\begin{align}
\label{eqn:NegTriAB}
 \mathcal{N}_{A(B)} &= \mathcal{N}_{A(C)} = \mathcal{N}_{B(C)}
 = \mathcal{N}_{B(A)} =    \mathcal{N}_{C(A)} = \mathcal{N}_{C(B)} \nn\\
 &= \max\big[ 0, \abs{X} -P \big]
\end{align}
yielding  
\begin{align}
\label{eqn:FinalPiTangleTriangle}
    \pi 
    &=
        \max
        \Bigg[0, \frac{\sqrt{C^2+8\abs{X}^2}}{2} - \frac{C}{2} -P \Bigg]^2  \nn\\ 
        &\quad -2\max\big[ 0, \abs{X} -P \big]^2
\end{align}
for the $\pi$-tangle \eqref{eqn:PiTangle}.

\begin{figure}[t]
\includegraphics[width=\linewidth]{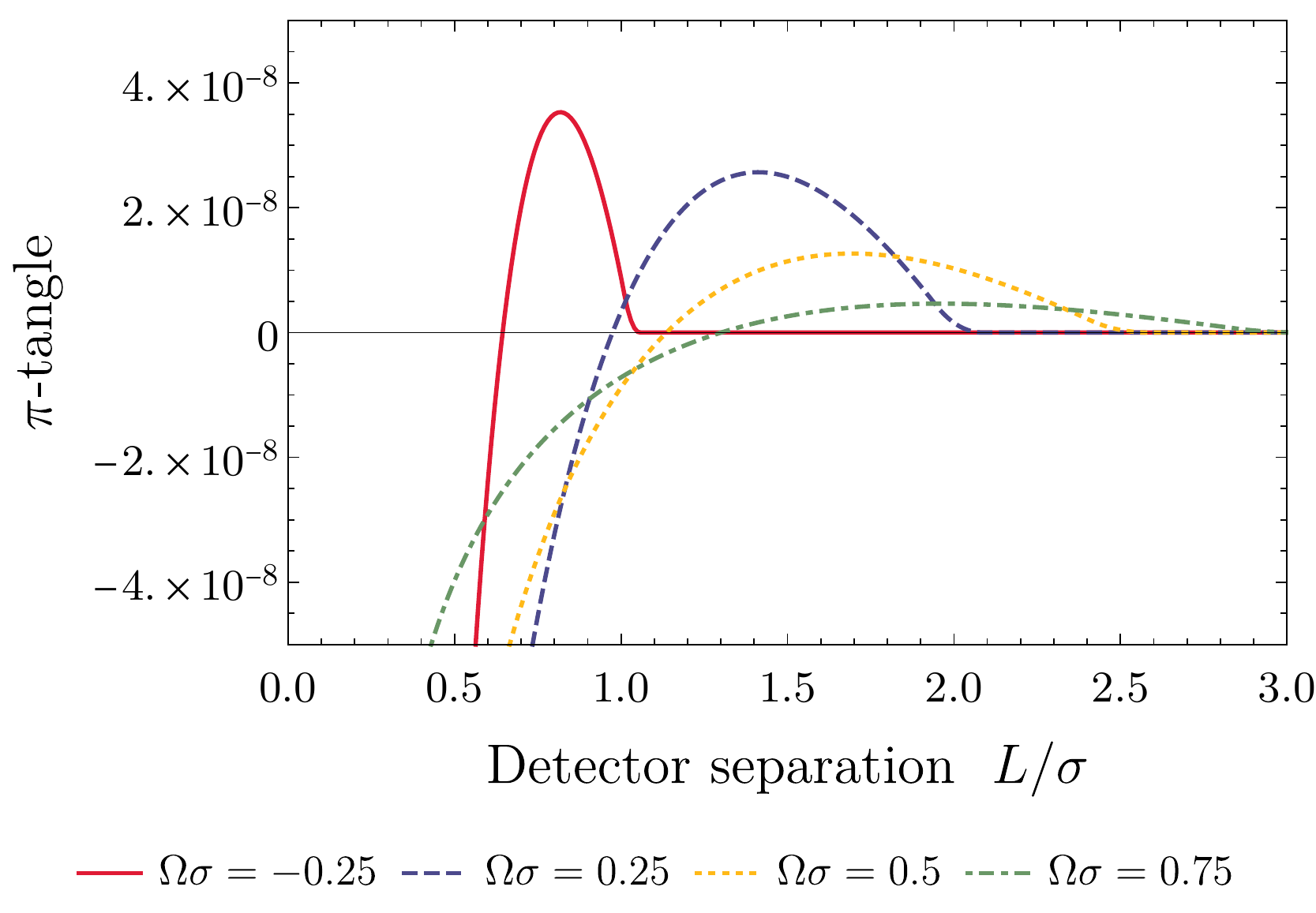}
\label{fig:2DTriangle}
\caption{~The $\pi$-tangle of the detectors in the triangle configuration with length $L/\sigma$ for various values of the energy gap, $\Omega$.  As the energy gap increases, the $\pi$-tangle reaches a lower maximum value but remains positive for larger values of detector separation.
The coupling constant  $\lambda=0.1$.
}
\label{fig:Zoom_2D_triangle}
\end{figure}

We plot the $\pi$-tangle from  \eqref{eqn:FinalPiTangleTriangle}
as a function  of the energy gap, $\Omega\sigma$, and the detector separation, $L/\sigma$ (the side of the triangle), in Fig.~\ref{fig:Triangle}. 
We observe that the $\pi$-tangle --- and hence the harvested tripartite entanglement --- is the largest for larger detector separations, contrary to intuitive expectation. 
This result also differs from the bipartite case \cite{pozas2015harvesting} where more entanglement is harvested for nearby detectors.
We also find that as the energy gap increases, tripartite entanglement can be harvested over broader ranges of larger detector separation, albeit in decreasing amounts. 
We illustrate this in Fig.~\ref{fig:Zoom_2D_triangle}, where we plot the $\pi$-tangle as a function
of detector separation for three different values of $\Omega \sigma$ --- these are different constant-$\Om$ cross-sections of Fig.~\ref{fig:Triangle}. We see that the harvested tripartite entanglement extends to larger values of $L/\sigma$ as $\Om \sigma$ increases.

One notable feature in Figs.~\ref{fig:Triangle} and~\ref{fig:Zoom_2D_triangle} is that there are regions in parameter space where the $\pi$-tangle becomes negative.  This would appear to violate the generalized  Coffman- Kundu-Wootters (CKW) inequality  for tripartite states \cite{Ou2007}. However the actual inequality is
\begin{equation}\label{genCKW}
    \mathcal{N}^2_{AB}+\mathcal{N}^2_{AC}\leq \textrm{min}\left[\mathcal{N}^2_{A(BC)}\right],
\end{equation}
where the minimization is over the possible pure-state decompositions \cite{Coffman:1999jd} of the 3-qubit mixed state
given by the density matrix $\rho_{ABC}$
in \eqref{eqn:ReducedDensMatrixTriangle}.
Since the negativity is a convex function \cite{Vidal2002negativity}, the minimum negativity $\textrm{min} [\mathcal{N}^2_{A(BC)} ]$ over a pure state decomposition is greater than or equal to the negativity $\mathcal{N}_{A(BC)}$ (and its counterparts) of the mixed state itself,  whenever we obtain $\pi > 0$  from  \eqref{eqn:PiTangle}. In this case,~\eqref{genCKW} is satisfied and the detectors have harvested tripartite entanglement.  However, if the $\pi$-tangle is negative, then the harvesting of tripartite entanglement is not guaranteed, since the inequality \eqref{genCKW} may or may not be saturated.
We discuss these issues further in  Appendix \ref{sec:Toy}.

\section{Linear  Arrangement}\label{sec:linear}

Another simple arrangement is the linear configuration --- the detectors are equally spaced along a line as shown in Fig.~\ref{2configs}(b). 
Although $P$ will be unchanged in this case, the matrix elements $X_{DD'}$ and $C_{DD'}$ will differ from the equilateral arrangement. 
They become
\begin{equation*}
    \begin{aligned}[c]
    X_{CB}
    =X_{BA}
    \equiv X_L
    \end{aligned}
    \qquad \text{and} \qquad 
    \begin{aligned}[c]
        C_{BC}=C_{AB} \equiv C_L
    \end{aligned}
\end{equation*}
for the two pairs of detectors at the same distance, $L\equiv L_{AB}=L_{BC}$, and
\begin{equation*}
    \begin{aligned}[c]
    X_{CA} \equiv X_{2L}
    \end{aligned}
    \qquad \text{and} \qquad
    \begin{aligned}[c]
        C_{AC} \equiv C_{2L}
    \end{aligned}
\end{equation*}
for the outermost pair of detectors. The density matrix in \eqref{eq:rhoABC2ndOrder} now has the form 
\begin{align}
\label{eqn:MatrixLinear}
    \rho_{ABC} 
    &=
        \begin{pmatrix}
        1-3P & 0 & 0 &  0 & X_{L}^* & X_{2L}^* & X_{L}^* & 0 \\
        0 & P & C_{L} & C_{2L} & 0 & 0 & 0 & 0 \\
        0 & C_{L} & P & C_{L} & 0 & 0 & 0 &0\\
        0 & C_{2L} & C_{L} & P & 0 & 0 & 0 & 0 \\
        X_{L} & 0 & 0 & 0 & 0  & 0 & 0 & 0 \\
        X_{2L} & 0 & 0 & 0 & 0 & 0 & 0 & 0 \\
        X_{L} & 0 & 0 & 0 & 0 & 0 & 0 & 0 \\
        0 & 0 & 0 & 0 & 0 & 0 & 0 & 0 \\
        \end{pmatrix} .
\end{align}

\begin{figure}[t]
    \centering
    \includegraphics[width=\linewidth]{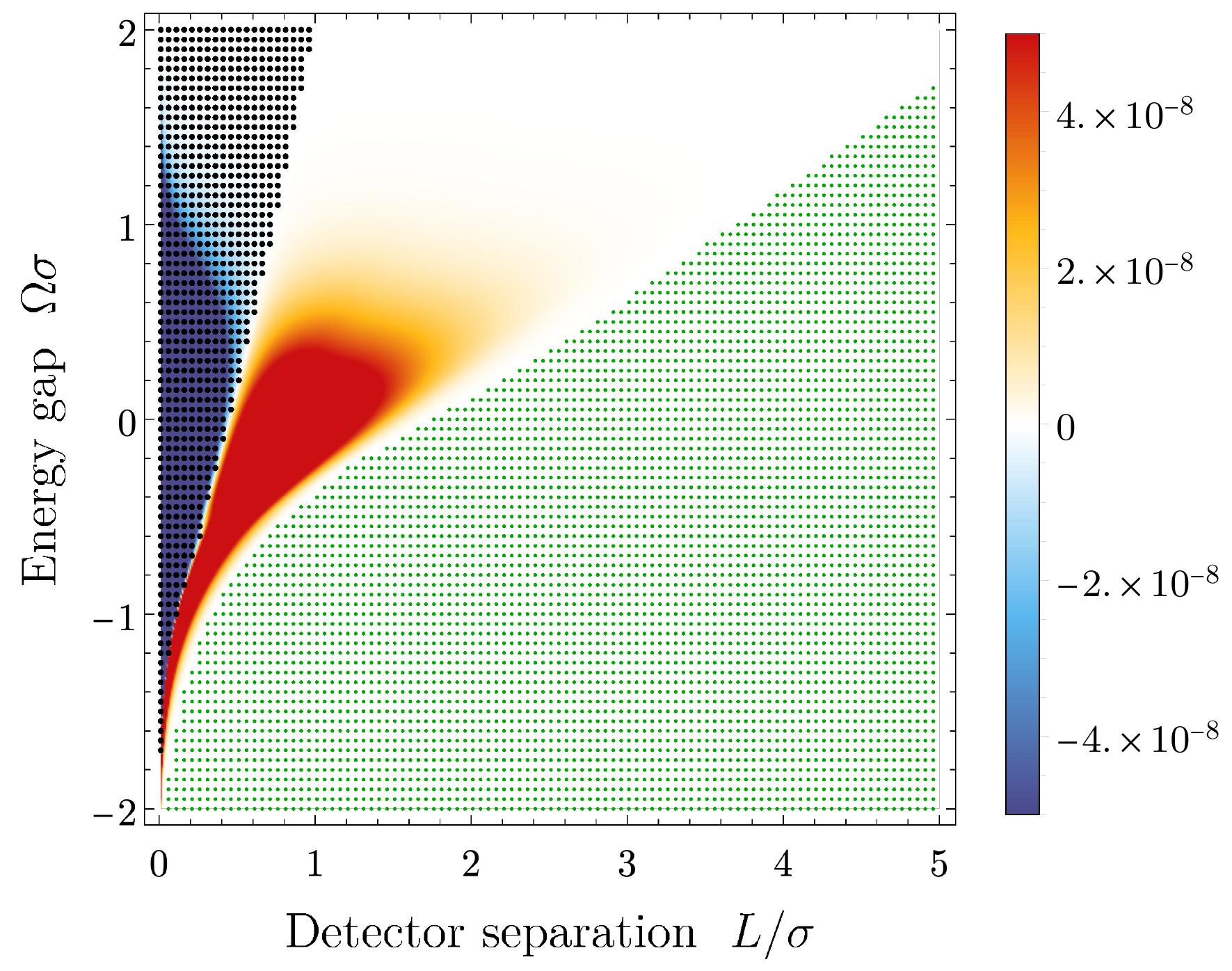}
    \caption{~The $\pi$-tangle as a function of $\Omega$ and detector separation $L$ for the linear configuration.  The total length of the line is $2L/\sigma$.
Negative values of $\Omega$ correspond to initially excited detectors.
    The coupling constant is set to $\lambda=0.1$.
    The green and black dotted regions represent zero and negative $\pi$-tangle, respectively.}
    \label{fig:Linear}
\end{figure}

We follow the same procedure as before to calculate the $\pi$-tangle. 
By symmetry, $\pi_A$ and $\pi_C$ will be the same, and we obtain
\begin{widetext}
\begin{align}
    \mathcal{N}_{A(BC)}=&\mathcal{N}_{C(AB)} = \max\Bigg[0, \frac{2}{\sqrt{3}}\sqrt{C_L^2+|X_L|^2+|X_{2L}|^2} \sin\left(\frac{\pi}{6}-\frac{1}{3}\arccos{\frac{3\sqrt{3}(C_LX_LX_{2L}^*+C_LX_L^*X_{2L})}{2(C_L^2+|X_L|^2+|X_{2L}|^2)^{3/2}}}\right) - P\Bigg] 
\end{align}
as well as 
\begin{align}
    \mathcal{N}_{A(B)} &= \mathcal{N}_{C(B)} =\max\big[ 0,\abs{X_L} - P\big], \\
    \mathcal{N}_{A(C)} &= \mathcal{N}_{C(A)} = \max\big[ 0,\abs{X_{2L}}-P\big] .
\end{align}
Hence 
\begin{align}
\label{eqn:LinearPiA}
    \pi_A =\pi_C  &=  \max\left[0,\frac{2}{\sqrt{3}}\sqrt{C_L^2+|X_L|^2+|X_{2L}|^2}\sin\left(\frac{\pi}{6}-\frac{1}{3}\arccos{\frac{3\sqrt{3}(C_LX_LX_{2L}^*+C_LX_L^*X_{2L})}{2(C_L^2+|X_L|^2+|X_{2L}|^2)^{3/2}}}\right)-P\right]^2\notag\\&\qquad - \max\big[ 0,\abs{X_L}-P\big]^2-\max\big[ 0,\abs{X_{2L}}-P\big]^2.
\end{align}
The computation of $\pi_B$ differs due to the relative difference in the detector  separations. We obtain
\begin{equation}
    \mathcal{N}_{B(AC)}
    =\max
    \left[ 0, 
            \frac{\sqrt{C_{2L}^2+8\abs{X_L}^2}}{2}
            - \frac{C_{2L}}{2} -  P 
    \right] 
    + \mathcal{O}(\lambda^4),
\end{equation}
yielding
\begin{align}
\label{eqn:LinearPiB}
    \pi_B  
    &=\max\left[0, \frac{\sqrt{C_{2L}^2+8\abs{X_L}^2}}{2}-\frac{C_{2L}}{2}-P\right]^2  -2 \max\big[0,\abs{X_L}-P\big]^2.
\end{align}
Finally, using  \eqref{eqn:LinearPiA} and   \eqref{eqn:LinearPiB}, we obtain
\begin{align}
\label{eqn:FinalLinearPi}
    \pi&=   
        \frac{2}{3}
         \max \left[0,\frac{2}{\sqrt{3}}\sqrt{C_L^2+|X_L|^2+|X_{2L}|^2}\sin\left(\frac{\pi}{6}-\frac{1}{3}\arccos{\frac{3\sqrt{3}(C_LX_LX_{2L}^*+C_LX_L^*X_{2L})}{2(C_L^2+|X_L|^2+|X_{2L}|^2)^{3/2}}}\right)-P\right]^2 \notag \\
    & \quad +\frac{1}{3}
    \max\left[0,\frac{\sqrt{C_{2L}^2+8\abs{X_L}^2}}{2}-\frac{C_{2L}}{2}-P\right]^2 
    -\frac{4}{3} \max\big[ 0,\abs{X_L}-P\big]^2 - \frac{2}{3}\max\big[ 0,\abs{X_{2L}}-P\big]^2.   
\end{align}
\end{widetext}

\begin{figure}[t]
\includegraphics[width=\linewidth]{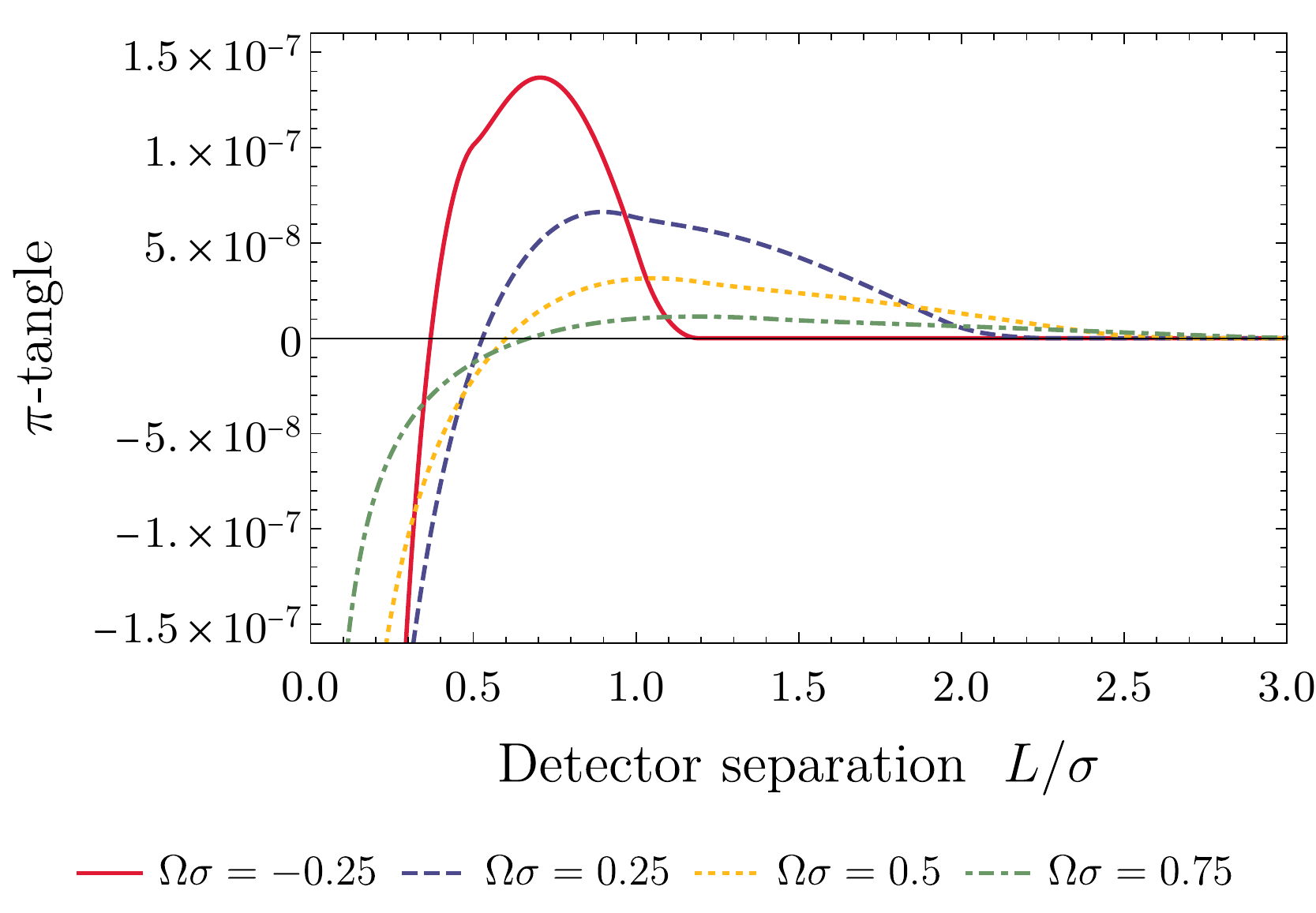}
\label{}
\caption{~The $\pi$-tangle of the detectors in the linear configuration with  total length $2L/\sigma$. 
 As the energy gap increases, the $\pi$-tangle attains diminishing maximal values, but remains positive for larger values of detector separation.
The coupling constant is set to $\lambda=0.1$.
}
\label{fig:Zoom_2D_line}
\end{figure}

In Fig.~\ref{fig:Linear} we depict the $\pi$-tangle from  \eqref{eqn:FinalLinearPi}
as a function  of the energy gap, $\Omega\sigma$, and the minimal detector separation, $L/\sigma$. 
We again see that tripartite entanglement is harvested a relatively larger separations than the bipartite case\cite{pozas2015harvesting}. %
 We also find that in the linear configuration we are able to obtain positive $\pi$-tangle for larger values of detector separation, at a fixed value of energy gap,  than in the equilateral triangle configuration.  This is made more explicit in Fig.~\ref{fig:Zoom_2D_line}, where we take cross-sections of Fig.~\ref{fig:Linear} at fixed values of $\Omega$ to plot the $\pi$-tangle as a function of detector separation.  By comparing Fig.~\ref{fig:Zoom_2D_line} with Fig.~\ref{fig:Zoom_2D_triangle}, we find for a given value of the energy gap, the $\pi$-tangle  reaches a larger maximum value in the linear case and remains positive over a larger range of detector separations.
These results indicate that it is more fruitful to harvest  entanglement from a linear arrangement as opposed to a triangle one for the same value of $L$ \cite{Avalos:2022oxr}.

\section{Scalene triangle}\label{sec:scalene}

As a generalization of the equilateral triangle and linear configuration, let us consider a scalene triangle arrangement  where the distance between any two  detectors is arbitrary.  As a result, the density matrix Eq.\eqref{eq:rhoABC2ndOrder} does not significantly simplify; we   leave the explicit calculations of the $\pi$-tangle to appendix \ref{sec:genpi}.

 In the previous two configurations, we considered the case where the distance between all three detectors increases. Here, however, we fix the distance between detectors $A$ and $C$ and take detector $B$ to be at different distances away from the other two. This is explicitly shown in Fig.~\ref{2configs}(a-ii).

\begin{figure}[t]
\includegraphics[width=\linewidth]{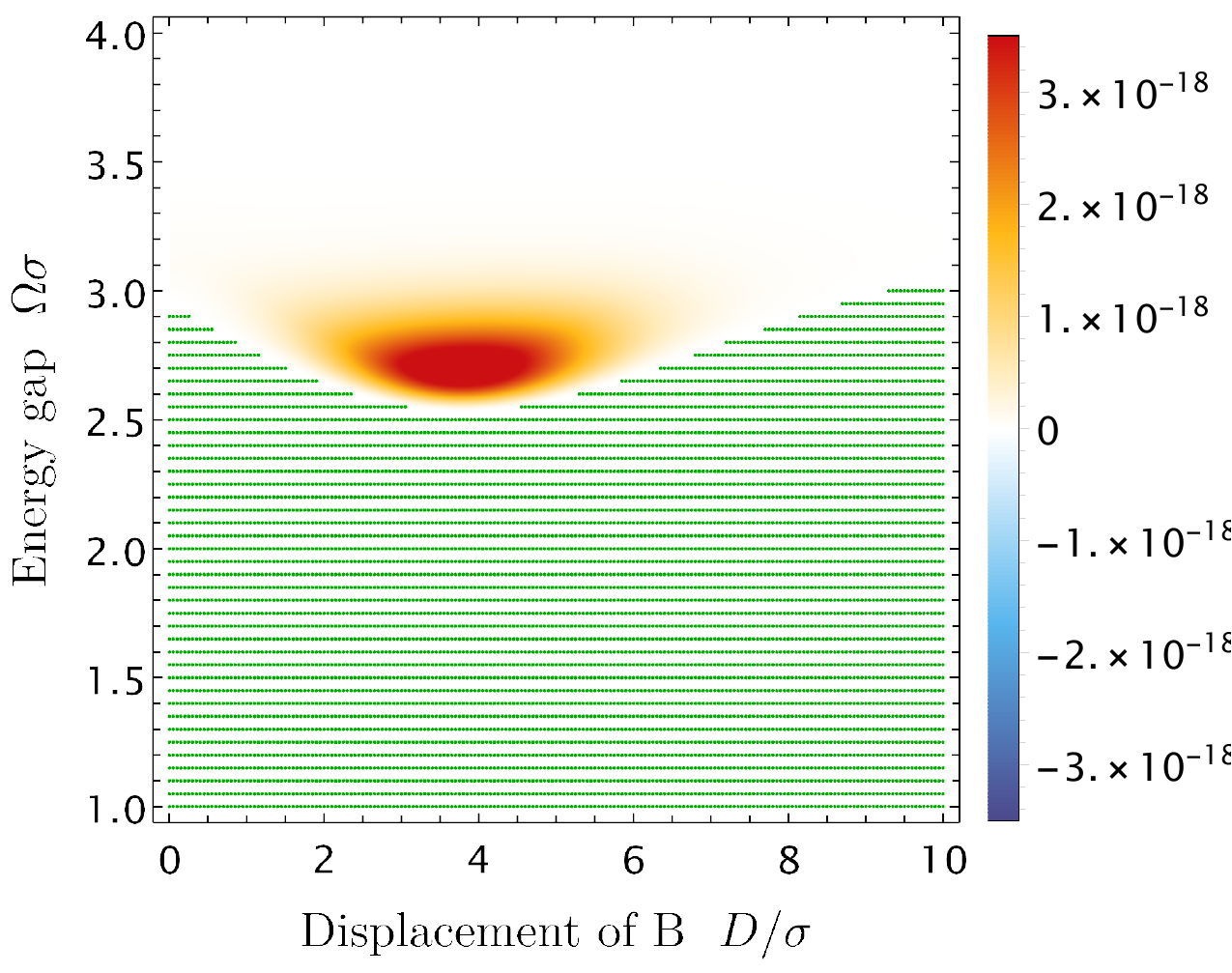}
\caption{~The $\pi$-tangle of the detectors in the scalene triangle configuration starting as an equilateral triangle with $L_{AC}=7\sigma$, for which all detectors are spacelike separated. 
 A displacement of $B$, $D/\sigma=0$ corresponds to an equilateral triangle. The green dotted region represents zero $\pi$-tangle. The coupling constant is set to $\lambda=0.1$.
}
\label{fig:Scalene}
\end{figure}

In Fig.~\ref{fig:Scalene}, we plot the $\pi$-tangle of the scalene triangular configuration as a function of the energy gap of the detectors and the displacement $L$ of $B$ along a line parallel to the line connecting detectors $A$ and $C$. 
A displacement of zero corresponds to an equilateral triangle.  The distance between detectors $A$ and $C$ is fixed to be $L_{AC}=7\sigma$, which was chosen using the criterion in \cite{pozas2015harvesting} to ensure that the detectors are spacelike separated throughout the interaction.  
Most striking, we find that unlike in the case of bipartite harvesting
\cite{pozas2015harvesting}, there will be tripartite entanglement (albeit a very small amount) following the interaction even when detector $B$ is far from the other two detectors. 

\begin{figure}[t]
\includegraphics[width=\linewidth]{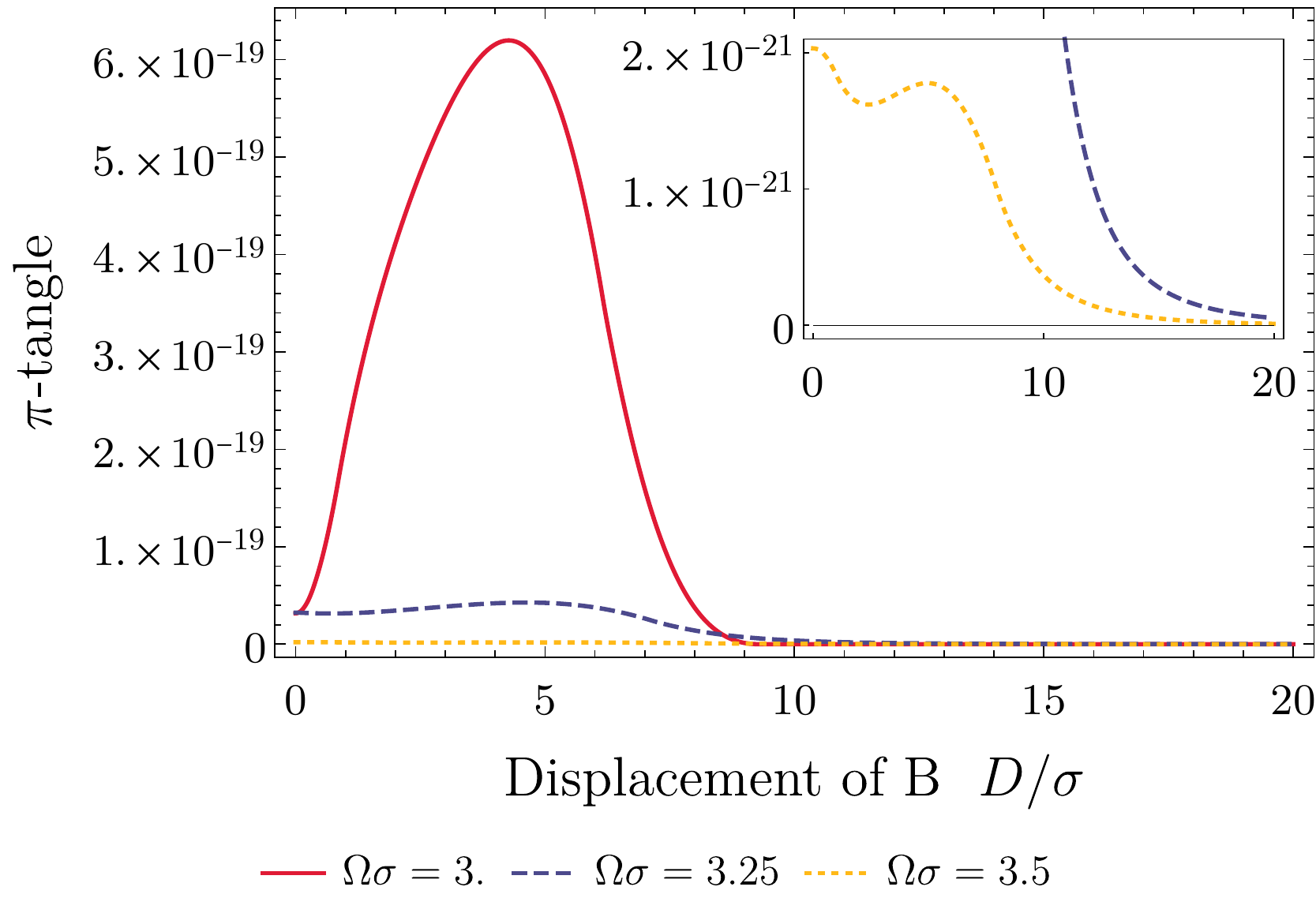}
\caption{~The $\pi$-tangle of the detectors in the scalene configuration with $L_{AC}=7\sigma$.
As the displacement of $B$ increases, the $\pi$-tangle approaches zero when $\Omega\sigma=3.25$ and $\Omega\sigma=3.5$. As the energy gap decreases, the $\pi$-tangle reaches a larger maximum value. However if the energy gap is too small, the $\pi$-tangle is zero for all displacements.
A displacement  $D/\sigma=0$ of  $B$ corresponds to an equilateral triangle. The inset depicts the $\Omega\sigma=3.25$ and $\Omega\sigma=3.5$ cases. 
The coupling constant is set to $\lambda=0.1$.
}
\label{fig:Zoom_2D_scalene}
\end{figure}

This is quite clear in Fig.~\ref{fig:Zoom_2D_scalene}, where we  plot  the $\pi$-tangle as a function of the displacement of detector $B$ at fixed values of the energy gap.   
We find that the $\pi$-tangle decreases as detector $B$ is moved away from the other two and approaches, but does not reach, zero. 

\begin{figure}[t]
\includegraphics[width=\linewidth]{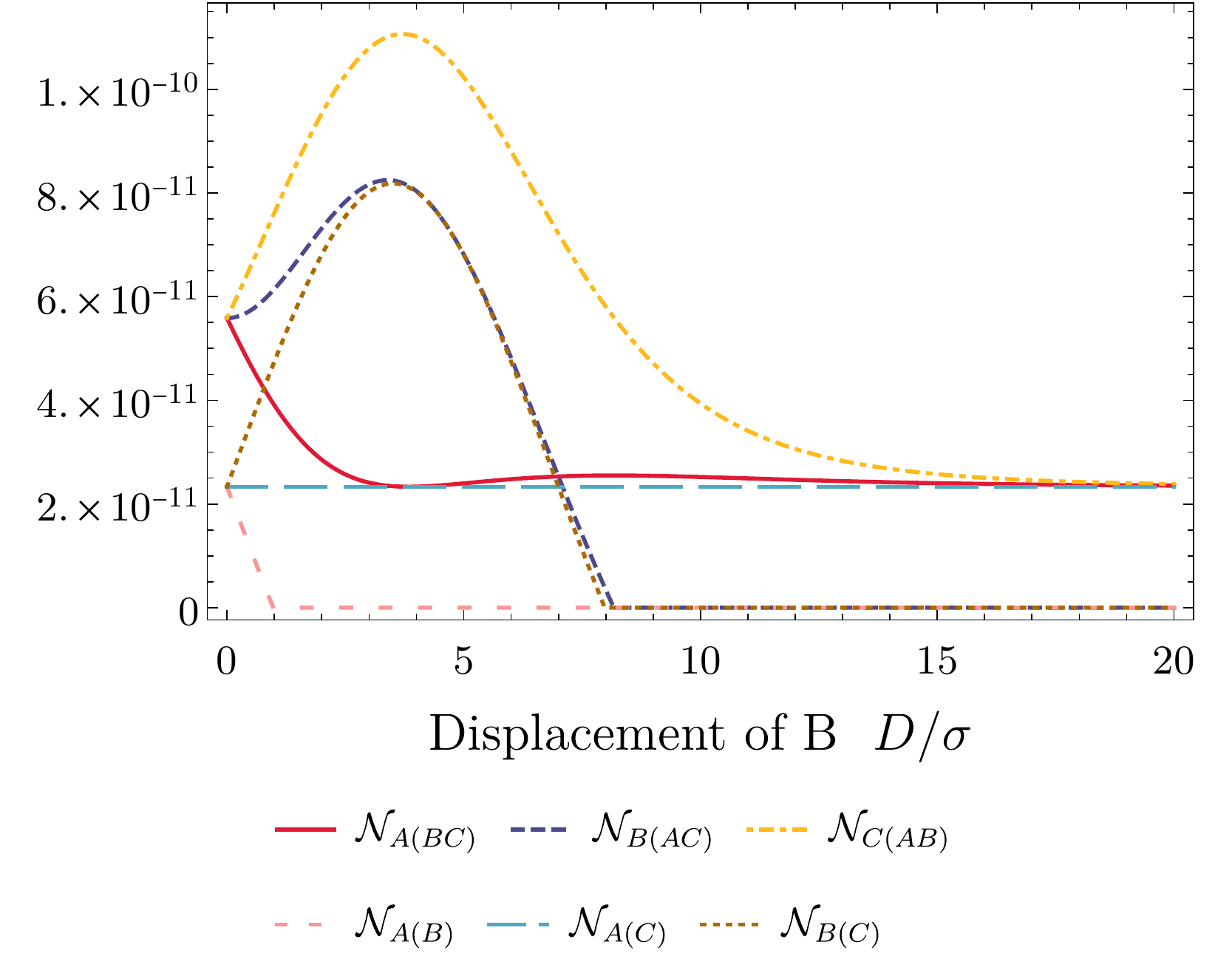}
\caption{~The tripartite and bipartite negativities of the detectors in the scalene configuration with $L_{AC}=7\sigma$.
As the displacement of $B$ increases, the bipartite negativities $\mathcal{N}_{A(B)}$ and $\mathcal{N}_{B(C)}$ and the tripartite negativity $\mathcal{N}_{B(AC)}$ go to zero, however the tripartite negativities $\mathcal{N}_{A(BC)}$ and $\mathcal{N}_{C(AB)}$ remain positive, leading to a positive $\pi$-tangle.
A displacement of $B=0$ corresponds to an equilateral triangle.  The energy gap of the detectors is $\Omega\sigma=3.5$ and the coupling constant is set to $\lambda=0.1$.
}
\label{fig:Negs_scalene}
\end{figure}

In Fig.~\ref{fig:Negs_scalene} we plot the bipartite and tripartite negativities as a function of the displacement of $B$. 
We find that for a fixed energy gap, once the displacement of $B$ is too large, i.e.\ the distance between detectors $A$ and $B$ and $C$ and $B$ respectively is too large, the bipartite negativities $\mathcal{N}_{A(B)}$ and $\mathcal{N}_{B(C)}$ become zero. 
In other words, when the $B$ is too far away, there is no bipartite negativity between detectors $A$ and $B$ and $C$ and $B$ respectively. 
Additionally, we find that the tripartite negativity between detector $B$ and the $(AC)$ subsystem, $\mathcal{N}_{B(AC)}$, also goes to zero at a similar displacement. 
However, we find that for the same energy gap, the tripartite negativities $\mathcal{N}_{A(BC)}$ and $\mathcal{N}_{C(AB)}$ remain positive for much larger displacements of $B$, yielding a non-zero $\pi$-tangle~\eqref{eqn:PiTangle}. 
Consequently it is possible to harvest tripartite entanglement for configurations where there is zero bipartite entanglement between some of the detectors. 
Furthermore, the tripartite negativities $\mathcal{N}_{A(BC)}$ and $\mathcal{N}_{C(AB)}$  appear to asymptote to the bipartite negativity, $\mathcal{N}_{A(C)}$ as the displacement of $B$ becomes very large, at which point the $\pi$-tangle will become zero.

 \section{Conclusion}
\label{sec:Conclusion}

As with the case for sharp switching
\cite{Avalos:2022oxr}, and in $(1+1)$ dimensions in the context of Gaussian Quantum Mechanics \cite{Lorek2014tripartite}, 
we find that it is actually easier to harvest tripartite 
entanglement than bipartite entanglement under certain circumstances.  In particular we find that tripartite entanglement can be harvested at comparatively larger detector separations and over broader ranges than the corresponding bipartite cases.    

The fact that the $\pi$-tangle becomes negative at sufficiently small detector separation $L$ for a given value of energy gap $\Om$ indicates that bipartite correlations overwhelm tripartite ones  at short distance. 
Proper minimization over pure-state decompositions will likely indicate positive values of the $\pi$-tangle persist to shorter distances. 
However, for any given $\Om$, we conjecture the inequality  \eqref{genCKW} to become saturated at some sufficiently small $L$, at which point tripartite harvesting is not possible. It would be interesting to see where this boundary is.

We also find, not surprisingly, dependence on the detector configuration. 
Although the general form of the density plots for the triangle and linear arrangements are similar, there are some notable distinctions. 
For a fixed value of detector energy gap, we find that there is a value of detector separation beyond which we cannot harvest entanglement in the linear arrangement; however, this value is larger than the equilateral triangular configuration case.  
In both the equilateral triangular and linear configurations, a sufficiently large   energy gap admits  harvesting of tripartite entanglement over arbitrary large detector separations, including spacelike separations, albeit in ever diminishing amounts. 

Finally, we find that if the distance between two detectors is fixed, there are  values of the energy gap of the detector where the $\pi$-tangle remains positive even if the third detector is far from the other two. 
Unlike   the equilateral triangle or linear configurations, the   energy gap (if sufficiently large) does not need to increase  to get positive $\pi$-tangle as the separation of the third detector is increased. 
This is true even when the three detectors are spacelike separated, and demonstrates that tripartite entanglement harvesting is possible for detector configurations that do not allow for bipartite entanglement harvesting.

Our results open up a number of interesting avenues of research.  Clearly all previous problems considering bipartite harvesting can be generalized to the tripartite case using the methods we have developed.
Interesting examples include   tripartite harvesting near moving mirrors, in curved spacetime, near a black hole, and across an  event horizon as one detector falls in. The range of different geometries for three detectors afford more possibilities for detector communication, and it would be interesting to see how such effects are manifest. And finally, a generalization to $n$-partite entanglement harvesting would be of considerable interest, though finding the right measure of entanglement would pose a challenge.\\

\section*{Acknowledgments}
All the numerical calculations and the figures were made via  Mathematica software \cite{Mathematica}. This work was supported in part by the Natural Sciences and Engineering Research Council of Canada. We are grateful to Meenu Kumari for helpful discussions.

\appendix

\begin{widetext}

\section{The $\pi$-tangle for the scalene triangle configuration}\label{sec:genpi}

Here we present the calculations for the $\pi$-tangle when the three detectors are in a scalene triangular configuration.

When the distance between each pair of detectors is arbitrary, the density matrix Eq.~\eqref{eq:rhoABC2ndOrder} becomes
\begin{align}
 {\rho}_{ABC} = \pmx{
		1-3P & 0 & 0 & 0 & X_{BC}^*  & X_{AC}^* & X_{AB}^* & 0 \\
		0 & P & C_{BC}^*  & C_{AC}^* & 0 & 0 & 0 & 0 \\
		0 & C_{BC} & P & C_{AB}^*  & 0  & 0 & 0 & 0 \\
		0 & C_{AC} & C_{AB}  & P  & 0 & 0 & 0 & 0 \\
		X_{BC} & 0 & 0 & 0 & 0 & 0 & 0 & 0 \\
		X_{AC} & 0 & 0 & 0 & 0 & 0 & 0 & 0 \\ 
		X_{AB} & 0 & 0 & 0 & 0 & 0 & 0 & 0 \\
		0 & 0 & 0 & 0 & 0 & 0 & 0 & 0
	}.
	\label{eq:MatrixScalene}
\end{align}

Following the same procedure   for the equilateral triangular and linear configurations, we obtain
\begin{align}
    \mathcal{N}_{A(BC)} &= \max\lb\{0,\frac{2}{\sqrt{3}}\sqrt{C_{BC}+|X_{AB}|^2+|X_{AC}|^2}\cos\lb[\frac{\pi}{3}+\frac{1}{3}\arccos\lb(\frac{3\sqrt{3}C_{BC}\lb(X_{AB}X_{AC}^*+X_{AB}^*X_{AC}\rb)}{2\big(C_{BC}^2+|X_{AB}|^2+|X_{AC}|^2\big)^{3/2}}\rb)\rb]-P\rb\} \nn\\
    &\quad + \max\lb\{0,\frac{2}{\sqrt{3}}\sqrt{C_{BC}+|X_{AB}|^2+|X_{AC}|^2}\sin\lb[\frac{\pi}{6}+\frac{1}{3}\arccos\lb(\frac{3\sqrt{3}C_{BC}\lb(X_{AB}X_{AC}^*+X_{AB}^*X_{AC}\rb)}{2\big(C_{BC}^2+|X_{AB}|^2+|X_{AC}|^2\big)^{3/2}}\rb)\rb]-P\rb\}\\
    \nn\\
    \mathcal{N}_{B(AC)} &= \max\lb\{0,\frac{2}{\sqrt{3}}\sqrt{C_{AC}+|X_{AB}|^2+|X_{BC}|^2}\cos\lb[\frac{\pi}{3}+\frac{1}{3}\arccos\lb(\frac{3\sqrt{3}C_{AC}\lb(X_{AB}X_{BC}^*+X_{AB}^*X_{BC}\rb)}{2\big(C_{AC}^2+|X_{AB}|^2+|X_{BC}|^2\big)^{3/2}}\rb)\rb]-P\rb\} \nn\\
    &\quad + \max\lb\{0,\frac{2}{\sqrt{3}}\sqrt{C_{AC}+|X_{AB}|^2+|X_{BC}|^2}\sin\lb[\frac{\pi}{6}+\frac{1}{3}\arccos\lb(\frac{3\sqrt{3}C_{AC}\lb(X_{AB}X_{BC}^*+X_{AB}^*X_{BC}\rb)}{2\big(C_{AC}^2+|X_{AB}|^2+|X_{BC}|^2\big)^{3/2}}\rb)\rb]-P\rb\}\\
    \nn\\
    \mathcal{N}_{C(AB)} &= \max\lb\{0,\frac{2}{\sqrt{3}}\sqrt{C_{AB}+|X_{AC}|^2+|X_{BC}|^2}\cos\lb[\frac{\pi}{3}+\frac{1}{3}\arccos\lb(\frac{3\sqrt{3}C_{AB}\lb(X_{AC}X_{BC}^*+X_{AC}^*X_{BC}\rb)}{2\big(C_{AB}^2+|X_{AC}|^2+|X_{BC}|^2\big)^{3/2}}\rb)\rb]-P\rb\} \nn\\
    &\quad + \max\lb\{0,\frac{2}{\sqrt{3}}\sqrt{C_{AB}+|X_{AC}|^2+|X_{BC}|^2}\sin\lb[\frac{\pi}{6}+\frac{1}{3}\arccos\lb(\frac{3\sqrt{3}C_{AB}\lb(X_{AC}X_{BC}^*+X_{AC}^*X_{BC}\rb)}{2\big(C_{AB}^2+|X_{AC}|^2+|X_{BC}|^2\big)^{3/2}}\rb)\rb]-P\rb\}
\end{align}
and
\begin{align}
    \mathcal{N}_{A(B)} &= \mathcal{N}_{B(A)} = \max\big[0,|X_{AB}|-P\big]\,, \\
    \mathcal{N}_{A(C)} &= \mathcal{N}_{C(A)} = \max\big[0,|X_{AC}|-P\big]\,, \\
    \mathcal{N}_{B(C)} &= \mathcal{N}_{C(B)} = \max\big[0,|X_{BC}|-P\big]\,.
\end{align}
The $\pi$-tangle is then easily computed using Eqs.~\eqref{eq:SubPi} and \eqref{eqn:PiTangle}.

\section{A toy model}\label{sec:Toy}

The introduction of the $\pi$-tangle \cite{Ou2007}
has generally been thought to satisfy the CKW inequality. Specficially
\textit{``For any pure $2\otimes2\otimes2$ states $\ket{\phi}_{ABC}$, the entanglement
quantified by the negativity between A and B, between A and C, and between A and the single object BC satisfies the following CKW-inequality-like monogamy inequality:}
\begin{equation*}
    \mathcal{N}^2_{AB}+\mathcal{N}^2_{AC}\leq\mathcal{N}^2_{A(BC)},
\end{equation*}
\textit{where $\mathcal{N}_{AB}$ and $\mathcal{N}_{AC}$ are the negativities of the mixed states''}. Although the  preceding inequality holds only for pure tripartite systems, the form of the preceding statement suggests that
 violations of the CKW inequality are perhaps unexpected and somewhat counter-intuitive.  To illustrate that such violations are not a consequence of 
 the perturbative expression given in \eqref{eq:rhoABC2ndOrder},
 we present a simple density matrix for a tripartite system of qubits that  can have negative $\pi$-tangle over a wide range of parameters.

Assume that we have a density matrix of the form
\begin{equation}
	\rho_{ABC} = \begin{pmatrix}
		1-3P-3E-\Sigma & 0 & 0 & 0 & X^* & X^* & X^* & 0 \\
		0 & P & C & C & 0 & 0 & 0 & 0 \\
		0 & C & P & C & 0 & 0 & 0 & 0\\
		0 & C & C & P & 0 & 0 & 0 & 0\\
		X & 0 & 0 & 0 & E & 0 & 0 & 0\\
		X & 0 & 0 & 0 & 0 & E & 0 & 0\\
		X & 0 & 0 & 0 & 0 & 0 & E & 0\\
		0 & 0 & 0 & 0 & 0 & 0 & 0 & \Sigma
		
	\end{pmatrix}\,,
	\label{eq:Toy}
\end{equation}
where $P,C,E,\Sigma \in \mathbb{R}$, which is clearly trace 1 and Hermitian.  
In order for this matrix to be a valid density matrix, it must have positive eigenvalues, which puts the following constraints on the elements of $\rho_{ABC}$:
\begin{subequations}
\begin{gather}
	E \ge 0\,, \\
	\Si \ge 0\,, \\
	P \ge C\,,  \label{eq:con3} \\
	P \ge -2C\,, \label{eq:con4} \\
	(1-3P-2E-\Sigma) \pm \sqrt{1-8E-16E^2-6P+24EP+9P^2+12|X|^2-2\Sigma+8E\Sigma+6P\Sigma+\Sigma^2} \ge 0\,. \label{eq:con5}
\end{gather}
\end{subequations}

From constraints Eqs.~\eqref{eq:con3} and \eqref{eq:con4}, we get the additional constraint:
\begin{equation}
	P \ge 0. \label{eq:con6}
\end{equation}
If we assume that
\begin{align}
	&\xi \coloneqq 1-3P-2E-\Sigma \ge 0\,, \label{eq:Ass1}
\end{align}
which will be true if $P,E,\Sigma\ll 1$, 
then constraint \eqref{eq:con5} will be satisfied provided:
\begin{gather}
	 (1-3P-2E-\Sigma)^2 \ge \big(1-8E-16E^2-6P+24EP+9P^2+12|X|^2-2\Sigma+8E\Sigma+6P\Si+\Sigma^2\big) \nn\\
	 \implies E\xi \ge E^2 + 3|X|^2\,, \label{eq:con5new}
\end{gather} 
which is consistent with the assumption \eqref{eq:Ass1} since $E\ge0$ and $|X|\ge0$.

The partial transpose of $\rho_{ABC}$ with respect to $A$ is
\begin{equation}
	\rho_{ABC}^{T_A} = \begin{pmatrix}
		1-3P-3E-\Sigma & 0 & 0 & 0 & X^* & C & C & 0 \\
		0 & P & C & X^* & 0 & 0 & 0 & 0 \\
		0 & C & P & X^* & 0 & 0 & 0 & 0\\
		0 & X & X & P & 0 & 0 & 0 & 0\\
		X & 0 & 0 & 0 & E & 0 & 0 & 0\\
		C & 0 & 0 & 0 & 0 & E & 0 & 0\\
		C & 0 & 0 & 0 & 0 & 0 & E & 0\\
		0 & 0 & 0 & 0 & 0 & 0 & 0 & \Sigma
		
	\end{pmatrix}\,,
\end{equation}
which only has two eigenvalues that can be negative while maintaining non-negative eigenvalues for $\rho_{ABC}$: 
\begin{align}
	& \frac{1}{2}\lb(2P+C - \sqrt{C^2+8|X|^2}\rb)\,, \label{eq:e4} \\
	& \frac{1}{2}
	\Bigg(
    	(1-3P-2E-\Sigma)  
    	- 
    	\sqrt{(1-3P-2E-\Sigma)^2+4\lb(2C^2-E+3E^2+EP+|X|^2+E\Sigma\rb)}
	\Bigg)\,. \label{eq:e7}
\end{align}
In particular eigenvalue \eqref{eq:e4} will only be non-negative if
\begin{equation}
	(2P+C)^2 \ge \big(C^2+8|X|^2\big) \implies P(P+C) \ge 2|X|^2 \label{eq:e4con}
\end{equation}
and eigenvalue \eqref{eq:e7} only be non-negative if
\begin{gather}
	(1-3P-2E-\Sigma)^2 \ge \big[(1-3P-2E-\Sigma)^2+4\lb(2C^2-E+3E^2+EP+|X|^2+E\Sigma\rb)\big] \nn\\
	\implies E\xi \ge E^2 + |X|^2 + 2C^2.  \label{eq:e7con}
\end{gather}
Note that provided $C^2>|X|^2$, eigenvalue \eqref{eq:e7} can be negative without contradicting condition \eqref{eq:con5new}.

Finally, we calculate the negativity of $\rho_{AB}=\Tr_C[\rho_{ABC}]$.
The partial transpose of $\rho_{AB}$ is
\begin{equation}
	\rh_{AB}^{T_A} = \begin{pmatrix}
		1-2P-3E-\Sigma & 0 & 0 & C \\
		0 & P+E & X^* & 0 \\
		0 & X & P+E & 0 \\
		C & 0 & 0 & E+\Sigma
	\end{pmatrix}\,,
\end{equation}
which again only has two eigenvalues that can be negative while maintaining non-negative eigenvalues for $\rho_{AB}$: 
\begin{align}
	&E+P-|X|\,, \label{eq:f1} \\
	&\frac{1}{2}\lb((1-2P-2E) - \sqrt{4C^2+(1-4E-2P-2\Sigma)^2}\rb)\,. \label{eq:f3}
\end{align}
Eigenvalue \eqref{eq:f1} will be non-negative if
\begin{equation}
	P+E > |X|
\end{equation}
and eigenvalue \eqref{eq:f3} will be non-negative if 
\begin{gather}
	(1-2P-2E)^2 \ge \big(4C^2+(-1+4E+2P+2\Sigma)^2\big) \nn\\
	\implies E\xi + \Sigma\xi -C^2 - E^2 - E\Si + PE + P\Sigma \ge 0.
	\label{eq:f3con}
\end{gather}

Under the assumption that
\begin{equation}
	P\ge\Sigma\,,
	\label{eq:Ass2}
\end{equation}
then condition \eqref{eq:f3con} can be related to condition \eqref{eq:e7con} by noticing that 
\begin{align}
	E\xi + \Sigma\xi -C^2 - E^2 - E\Sigma + PE + P\Sigma 
	&= 
	    \lb(E\xi-E^2-|X|^2-2C^2\rb) 
	    + |X|^2 + C^2 + \Sigma\xi + P\Sigma + E(P-\Sigma) \nn\\
	&\ge 
	    \lb(E\xi-E^2-|X|^2-2C^2\rb)\,,
\end{align}
meaning that if Eq.~\eqref{eq:f3con} is not satisfied then neither will Eq.~\eqref{eq:e7con};  if Eq.~\eqref{eq:e7con} is satisfied then Eq.~\eqref{eq:f3con} will be too.

The only way that the CKW inequality is violated will be if $\mathcal{N}_{A(B)}$ is nonzero. In other words,   at least one of the eigenvalues of the partial transpose of $\rho_{AB}$ must be negative. We will first consider the case where eigenvalue \eqref{eq:f1} is negative, followed by   the case  where ~\eqref{eq:f3} is negative. 
We will also show that under the assumptions that \eqref{eq:Ass1} and \eqref{eq:Ass2} are both valid, it is not possible for both eigenvalues to be simultaneously negative.

\subsection*{Case \#1: Assume that eigenvalue \eqref{eq:f1} is negative} 

If eigenvalue \eqref{eq:f1} is negative, then $|X|>P+E$ and we also have $|X|^2>C^2$ (since $P+E\ge P\ge C$). 
This means that in order that $\rho_{ABC}$ remain a valid density matrix, eigenvalue \eqref{eq:e7} must be non-negative, implying that eigenvalue \eqref{eq:f3} must be also non-negative.
Additionally eigenvalue \eqref{eq:e4} must also be negative since
\begin{equation}
	|X|>P+E\ge P \implies 2|X|^2 > 2P^2 = P^2+P^2 \ge P^2+PC = P(P+C)\,.
\end{equation}

In this case, the $\pi$-tangle is
\begin{align}
	\pi &= \mathcal{N}_{A(BC)}^2 - 2\mathcal{N}_{A(B)}^2 \nn\\
	&= \Bigg(\frac{1}{2}\lb(\sqrt{C^2+8|X|^2}-2P-C\rb)\Bigg)^2 - 2\big(|X|-P-E\big)^2\,, \label{eq:Case1}
\end{align}
which can be simplified by considering a Taylor expansion of the matrix elements of $\rho_{ABC}$ in powers of the coupling strength, $\lambda\ll1$:
\begin{align}
	P &= \lambda^2 P_2 + \lambda^4 P_4 + \lambda^6 P_6 + \Or\lb(\lambda^8\rb)\,, \nn\\
	C &= \lambda^2 C_2 + \lambda^4 C_4 + \lambda^6 C_6 + \Or\lb(\lambda^8\rb)\,, \nn\\
	X &= \lambda^2 X_2 + \lambda^4 X_4 + \lambda^6 X_6 + \Or\lb(\lambda^8\rb)\,, \label{eq:pert} \\
	E &= \lambda^4 E_4 + \lambda^6 E_6 + \Or\lb(\lambda^8\rb)\,, \nn\\
	\Si &= \lambda^6 \Si_6 + \Or\lb(\lambda^8\rb)\,. \notag 
\end{align}
Under this expansion, the $\pi$-tangle becomes:
\begin{align}
	\pi %
	&\approx \lambda^4\Bigg(\frac{1}{4}\lb(\sqrt{C_2^2+8|X_2|^2}-2P_2-C_2\rb)^2-2\big(|X_2|-P_2\big)^2\Bigg)\,, \label{eq:Case1pert}
\end{align}  
which will be non-negative if
\begin{gather}
	\Big(8P_2|X_2|+C_2^2+2P_2C_2-2P_2^2\Big)^2 \ge \lb((2P_2+C_2)\sqrt{C_2^2+8|X_2|^2}\rb)^2 \nn\\
	\iff 4\Big[P_2^2(P_2-C_2)^2-2\big(|X_2|-P_2\big)\Big(C_2^2\big(2|X_2|-P_2\big)-|X_2|(2P_2-C_2)^2\Big)\Big] \ge 0.
	\label{eq:Case1con}
\end{gather}

\begin{figure}[t]
	\centering
	\includegraphics[width=0.6\textwidth]{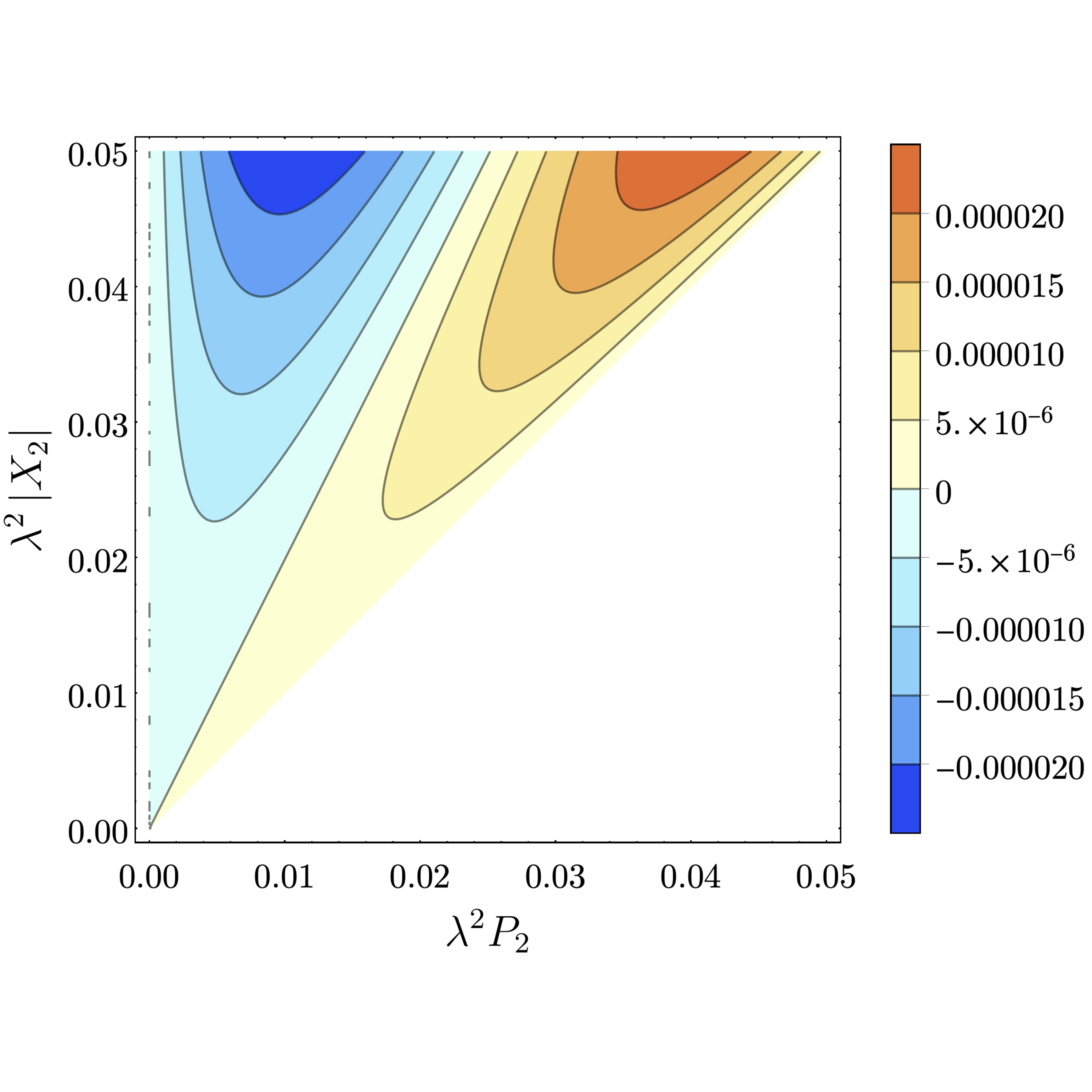}
	\caption{%
		\ A plot of the perturbative $\pi$-tangle \eqref{eq:Case1pert} where the value of $C_2$ is fixed to be $C_2=0.9P_2$ and $\la=0.1$.  There are regions of the parameter space where the $\pi$-tangle is negative.  The region where $P_2>|X_2|$ is excluded from the plot, since this corresponds to eigenvalue \eqref{eq:f1} being non-negative.
	}
	\label{fig:Case1}  
\end{figure}

As the condition  in~\eqref{eq:Case1con} is not particularly instructive, we plot the perturbative $\pi$-tangle \eqref{eq:Case1pert} in Fig.~\ref{fig:Case1}, where it is clear that there are regions of the parameter space where the $\pi$-tangle is negative, meaning the CKW inequality is not satisfied.  For example, if
\begin{equation*}
	\lambda=0.1,\ P_2=1,\ C_2=0.9,\ |X_2|=4 \implies \pi \approx \frac{3261-29\sqrt{12\ 881}}{2\ 000\ 000} \approx -1.52\times10^{-5}.
\end{equation*}

Additionally, if the non-perturbative expression for the $\pi$-tangle~\eqref{eq:Case1}  is considered, it is still possible to find regions of the parameter space where the CKW inequality is not satisfied.  For example, 
\begin{equation*}
	P=0.01,\ C_2=0.009,\ |X_2|=0.04,\ E=0.00011 \implies \pi = \frac{8\ 218\ 379 - 72\ 500\ \sqrt{12\ 881}}{5\ 000\ 000\ 000} \approx -1.99\times10^{-6}.
\end{equation*}

\subsection*{Case \#2: Assume that eigenvalue \eqref{eq:f3} is negative}

If eigenvalue \eqref{eq:f3} is negative, then $\sqrt{4C^2+(1-4E-2P-2\Sigma)^2}>(1-2P-2E)$ and eigenvalue $\eqref{eq:e7}$ will also be negative. 
Recall that in order to ensure that $\rho_{ABC}$ is a valid density matrix, we also require $C^2>|X|^2$. 
This also means that eigenvalue \eqref{eq:f1} is non-negative since $|X|^2 < C^2 \le P^2 \le (P+E)^2$.
Also notice that if $C\ge0$, then
\begin{equation*}
	P(P+C) 
	\ge 
	    C(C+C) 
    = 
        2C^2 
    >
        2|X|^2\,, %
\end{equation*}
and if $C<0$, then Eq.~\eqref{eq:con4} becomes $P\ge-2C=2|C|$, and
\begin{equation}
	P(P+C) = P\big(P-|C|\big) \ge 2|C|\big(2|C|-|C|\big) 
	= 2|C|^2 
	= 2C^2 > 2|X|^2\,,
\end{equation}
so, eigenvalue \eqref{eq:e4} is also non-negative regardless of the sign of $C$.
Therefore, the $\pi$-tangle is
\begin{align}
	\pi &= \mathcal{N}_{A(BC)}^2 - 2\mathcal{N}_{A(B)}^2 \nn\\
	&= \left[\frac{1}{2}\Big(\sqrt{(1-3P-2E-\Sigma)^2+4\lb(2C^2-E+3E^2+EP+|X|^2+E\Sigma\rb)}  - (1-3P-2E-\Sigma)\Big)\right]^2 \nn\\
	&\qquad\qquad -2 \Bigg[\frac{1}{2}\lb(\sqrt{4C^2+(1-4E-2P-2\Sigma)^2} - (1-2P-2E)\rb)\Bigg]^2\label{eq:Case2}\\
	&\approx \lambda^8\Big(2\big(C_2^2+|X_2|^2\big)^2-\big(E_4+|X_2|^2\big)^2\Big), \label{eq:Case2pert}
\end{align}
where the last line is from the perturbative expansion of $\rho_{ABC}$ [Eq.~\eqref{eq:pert}].  In this case, the $\pi$-tangle will be non-negative if
\begin{equation}
	E_4 \le \big(\sqrt{2}-1\big)|X_2|^2 + \sqrt{2}C_2^2\,.
\end{equation}
Once again, we can find regions of the parameter space where the $\pi$-tangle is negative, meaning the CKW inequality is not satisfied.  For example if
\begin{equation*}
	\lambda=0.1,\ P_2=1,\ C_2=0.9,\ |X_2|=0.8,\ E_4=1.5 \implies \pi = -\frac{1873}{500\ 000\ 000\ 000} \approx -3.75\times10^{-9}.
\end{equation*}

Additionally, if the non-perturbative expression for the $\pi$-tangle \eqref{eq:Case2} is considered, it is still possible to find regions of the parameter space where the CKW inequality is not satisfied.  For example if
\begin{gather*}
	P=0.01,\ C=0.009,\ |X|=0.008,\ E=0.00015,\ \Sigma= 10^{-6}\\
	\implies \pi = \frac{-978\ 879\ 246\ 503+35\ 269\ 200\ \sqrt{2\ 961\ 556\ 921}-969\ 699\ \sqrt{940\ 638\ 421\ 201}}{2\ 000\ 000\ 000\ 000} \approx -2.43\times10^{-9}.
\end{gather*}

\end{widetext}

\newpage

\bibliography{ref}

\end{document}